\documentclass[11pt,twoside]{article}
\usepackage{asp2010}

\resetcounters

\begin{document}
\title{Discovery of the closest hot subdwarf binary with white dwarf companion}
\author{S. Geier$^1$, T. R. Marsh$^2$, B. H. Dunlap$^3$, B. N. Barlow$^4$, V. Schaffenroth$^1$, E. Ziegerer$^1$, U. Heber$^1$, T. Kupfer$^{5}$, P. F. L. Maxted$^6$, B. Miszalski$^{7,8}$, A. Shporer$^{9,10,11}$, J. H. Telting$^{12}$, R. H. \O stensen$^{13}$, S. J. O'Toole$^{14}$, B. T. G\"ansicke$^2$, R. Napiwotzki$^{15}$
\affil{$^1$Dr. Karl Remeis-Observatory \& ECAP, Astronomical Institute,
Friedrich-Alexander University Erlangen-Nuremberg, Sternwartstr. 7, D~96049 Bamberg, Germany}
\affil{$^2$Department of Physics, University of Warwick, Conventry CV4 7AL, UK}
\affil{$^3$Department of Physics and Astronomy, University of North Carolina, Chapel Hill, NC 27599-3255, USA}
\affil{$^4$Department of Astronomy \& Astrophysics, The Pennsylvania State University, 525 Davey Lab, University Park, PA 16802, USA}
\affil{$^5$Department of Astrophysics/IMAPP, Radboud University Nijmegen, P.O. Box 9010, 6500 GL Nijmegen, The Netherlands}
\affil{$^6$Astrophysics Group, Keele University, Staffordshire, ST5 5BG, UK}
\affil{$^7$South African Astronomical Observatory, P.O. Box 9, Observatory, 7935, South Africa}
\affil{$^8$Southern African Large Telescope Foundation, P.O. Box 9, Observatory, 7935, South Africa}
\affil{$^9$Las Cumbres Observatory Global Telescope Network, 6740 Cortona Drive, Suite 102, Santa Barbara, CA 93117, USA}
\affil{$^{10}$Department of Physics, Broida Hall, University of California, Santa Barbara, CA 93106, USA}
\affil{$^{11}$Division of Geological and Planetary Sciences, California Institute of Technology, Pasadena, CA 91125, USA}
\affil{$^{12}$Nordic Optical Telescope, Apartado 474, 38700 Santa Cruz de La Palma, Spain}
\affil{$^{13}$Institute of Astronomy, K.U.Leuven, Celestijnenlaan 200D, B-3001 Heverlee, Belgium}
\affil{$^{14}$Australian Astronomical Observatory, PO Box 296, Epping, NSW, 1710, Australia}
\affil{$^{15}$Centre of Astrophysics Research, University of Hertfordshire, College Lane, Hatfield AL10 9AB, UK}}

\vspace{-0.5cm}

\begin{abstract}
We report the discovery of an extremely close, eclipsing binary system. A white dwarf is orbited by a core He-burning compact hot subdwarf star with a period as short as  $\simeq0.04987\,{\rm d}$ making this system the most compact hot subdwarf binary discovered so far. The subdwarf will start to transfer helium-rich material on short timescales of less than $50\,{\rm Myr}$. The ignition of He-burning at the surface may trigger carbon-burning in the core although the WD is less massive than the Chandrasekhar limit ($>0.74\,M_{\rm \odot}$) making this binary a possible progenitor candidate for a supernova type Ia event. 
\end{abstract}

\section{Introduction}

After finishing core-hydrogen-burning the progenitors of hot subdwarf stars (sdBs) leave the main sequence and evolve to red giants before igniting helium and settling down on the extreme horizontal branch. Unlike normal stars, the sdB progenitors must have experienced a phase of extensive mass loss on the red giant branch, in order to explain the high temperatures and gravities observed at the surface of hot subdwarf stars. After consumption of the helium fuel they evolve directly to white dwarfs (WDs) avoiding a second red-giant phase. What causes this extensive mass loss remains an open question.

About half of the sdB stars reside in close binaries with periods as short as $<0.1$\,d \citep{maxted01,napiwotzki04}. Because the components' separation in these systems is much less than the size of the subdwarf progenitor in its red-giant phase, these systems must have experienced a common-envelope and
spiral-in phase \citep{han02,han03}. Although most of the close companions to sdB stars are low-mass main sequence stars, brown dwarfs or low mass WDs ($\simeq0.5\,M_{\rm \odot}$), more massive compact companions like WDs, neutron stars or black holes have been either observed or predicted by theory \citep{geier11c,geier10a,geier10b}.

Subdwarf binaries with massive WD companions turned out to be candidates for supernova type Ia (SN Ia) progenitors because these systems lose angular momentum due to the emission of gravitational waves and shrink. Mass transfer or the subsequent merger of the system may cause the WD to reach the Chandrasekhar limit and explode as a SN~Ia \citep[see review by][]{wang12}. One of the best known candidate systems for the double degenerate merger scenario is the sdB+WD binary KPD\,1930$+$2752 \citep{maxted00,geier07}. 

\begin{figure*}[hp!]
\begin{center}
 \includegraphics[width=8cm]{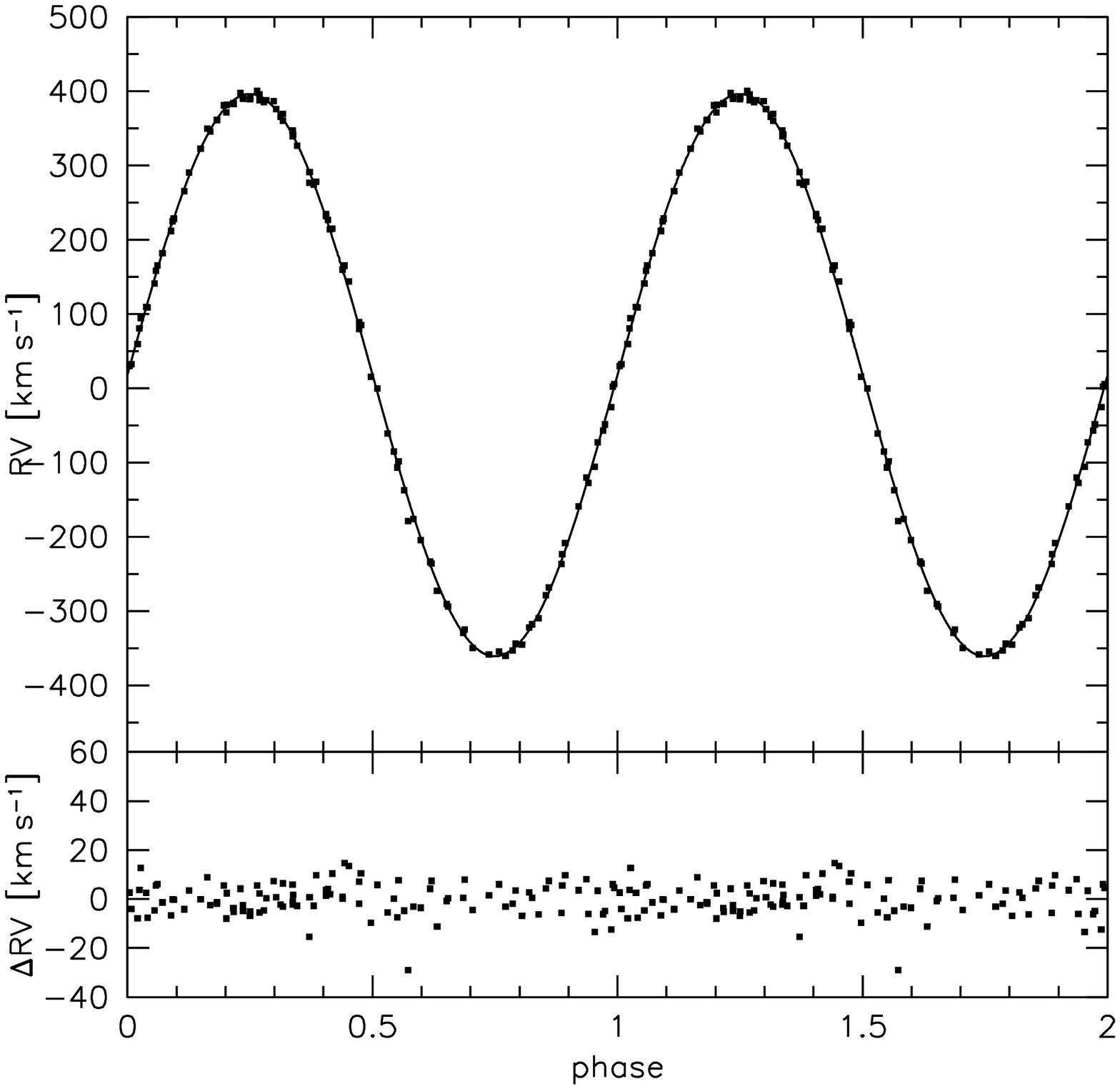}
 \caption{Radial velocity curve of CD$-$30\,11223 derived from 105 spectra taken with WHT/ISIS.}
 \label{rv}
\end{center}
\end{figure*}

\section{Observations}

The sdB+WD binary CD$-$30\,11223 ($\alpha_{2000}=14^{\rm h}11^{\rm m}16{\stackrel{\rm s}{\displaystyle 
.}}2$, $\delta_{2000}=-30^{\rm \circ}53'03''$, $m_{\rm V}=11.9\,{\rm mag}$) has been discovered in the course of the MUCHFUSS project, which aims at finding hot subdwarf binary systems with massive compact companions \citep{geier11a}. It was selected as an UV-excess object and spectroscopically identified to be an sdB star by \citet{vennes11}. 

The star was chosen as a bright backup target for our MUCHFUSS follow-up campaign. Due to bad observing conditions, which did not allow us to observe our main targets, two medium resolution spectra ($R\simeq2200,\lambda=4450-5110\,{\rm \AA}$) were taken consecutively with the EFOSC2 spectrograph mounted at the ESO\,NTT at June 10, 2012. The spectra showed a very high radial velocity shift ($\simeq600\,{\rm km\,s^{-1}}$).

The star was observed by the SuperWASP planetary transit survey \citep{pollacco06}. The light curve was downloaded from the public archive\footnote{http://www.wasp.le.ac.uk} and consists of more than $13\,000$ measurements taken from May 4, 2006 to August 9, 2008. The light curve shows a clear variation caused by the ellipsoidal deformation and Doppler boosting of the sdB primary \citep{geier07,bloemen11}, which allowed us to derive the orbital period based on fitting an harmonic series ($P=0.0489790717\pm0.0000000038\,{\rm d}$). The amplitude of the ellipsoidal variation is $\simeq4\%$ and therefore the highest ever measured in an sdB binary. 

First spectroscopic follow-up data was obtained with the grating spectrograph mounted on the SAAO-1.9m telescope at July 2, 2012. The RV-curve measured from 18 single spectra confirmed the short orbital period and a high RV-semiamplitude ($K=370\pm14\,{\rm km\,s^{-1}}$). In order to improve the orbital solution, we took another 105 spectra with the ISIS spectrograph ($R\simeq4000,\lambda=3440-5270\,{\rm \AA}, T_{\rm exp}=2\,{\rm min}$) mounted at the WHT during a dedicated MUCHFUSS-follow-up-run from July 9 to 12, 2012. Another 175 spectra were taken with the the Goodman spectrograph mounted at the SOAR telescope ($R\simeq7700, \lambda=3700-4400\,{\rm \AA}, T_{\rm exp}=1\,{\rm min}$) at July 16, 2012.

\begin{figure*}[hp!]
\begin{center}
 \includegraphics[width=8cm,angle=90]{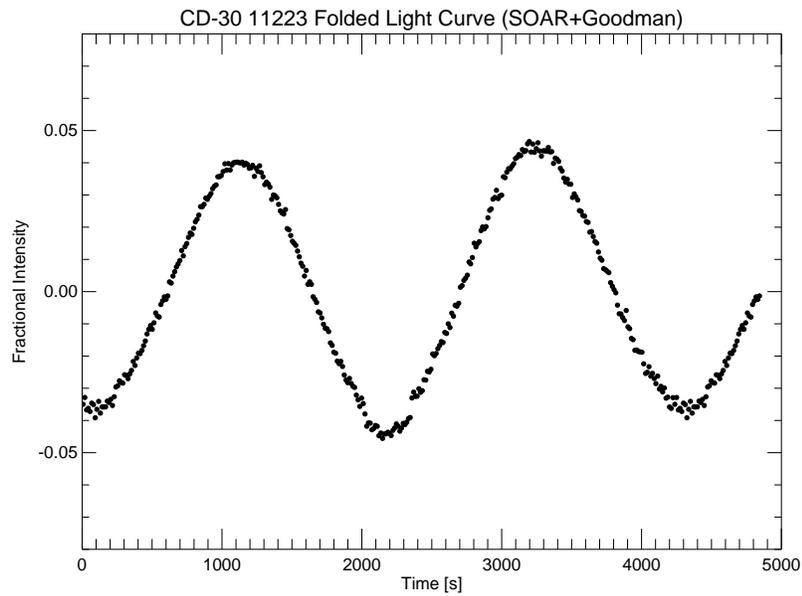}
 \caption{V-band light curve of CD$-$30\,11223 taken with SOAR/Goodman and folded to the orbital period.}
 \label{lc}
\end{center}
\end{figure*}

Additionally, $3.6\,{\rm hr}$ of time-series photometry in the V-band ($T_{\rm exp}=3\,{\rm s}$) were taken with SOAR/Goodman at July 6, 2012 under photometric conditions. Eclipses are detected in this high-quality light curve. The eclipse of the sdB by the WD companion is clearly visible and the flat bottom seen in the other minima indicates a shallow secondary eclipse as well (see Fig.~\ref{lc}).

\section{Spectral analysis}

The radial velocities (RVs) were measured by fitting a set of mathematical functions (Gaussians, Lorentzians and polynomials) mimicking the typical Voigt profile of spectral lines to the hydrogen Balmer lines and helium lines using the FITSB2 routine. The profiles are fitted to all suitable lines simultaneously using $\chi^{2}$-minimization and the RV-shift with respect to the rest wavelengths is measured. Assuming a circular orbit sine curves were fitted to the RV data points in fine steps over a range of test periods close to the orbital period derived from photometry. The two datasets obtained with ISIS and Goodman are treated separately to investigate systematic errors. Details about the analysis method and error estimation are given in \citet{geier11b}. 

The derived orbital parameters from the ISIS dataset ($K=378.6\pm1.0\,{\rm km\,s^{-1}}$, $\gamma=17.6\pm0.7\,{\rm km\,s^{-1}}$, see Fig.~\ref{rv}) and the Goodman dataset ($K=374.5\pm1.1\,{\rm km\,s^{-1}}$, $\gamma=21.3\pm0.8\,{\rm km\,s^{-1}}$) are roughly consistent. The deviation in system velocity is most likely caused by a slight systematic zero-point shift between the two instruments. CD$-$30\,11223 is the sdB binary with the shortest orbital period and the highest RV-amplitude discovered so far.

The atmospheric parameters effective temperature $T_{\rm eff}$, surface gravity $\log\,g$, helium abundance $\log\,y$ and projected rotational velocity were determined by fitting simultaneously the observed hydrogen and helium lines of the single spectra with metal-line-blanketed LTE model spectra as described in \citet{geier07}. No significant variations of the parameters with orbital phase have been detected. Average values and standard deviations have been calculated for the ISIS ($T_{\rm eff}=28800\pm200\,{\rm K}$,$\log{g}=5.67\pm0.03$,$\log{y}=-1.50\pm0.07$, $v_{\rm rot}\sin{i}=180\pm8\,{\rm km\,s^{-1}}$) and Goodman datasets ($T_{\rm eff}=29600\pm300\,{\rm K}$,$\log{g}=5.65\pm0.05$,$\log{y}=-1.46\pm0.14$, $v_{\rm rot}\sin{i}=174\pm12\,{\rm km\,s^{-1}}$), separately. The derived parameters are consistent with literature values within the uncertainties \citep{vennes11}.

\section{Deriving the mass of the unseen companion}

Since the spectra of the CD$-$30\,11223 are single-lined, they
contain no information about the orbital motion of the companion, and
thus only the mass function 
$$f_{\rm m} = \frac{M_{\rm comp}^3
\sin^3i}{(M_{\rm comp} + M_{\rm sdB})^2} = \frac{P K^3}{2 \pi G}$$ 
can be calculated. Although the RV semi-amplitude $K$ and the period $P$
are determined by the RV curve, $M_{\rm sdB}$, $M_{\rm comp}$ and the inclination angle 
$i$ remain free parameters. Assuming a reasonable value for
$M_{\rm sdB}$ only a lower limit can be derived for $M_{\rm comp}$.

If the rotation of the companion is synchronised to its orbital motion, which is the case in CD$-$30\,11223, 
the rotational velocity $v_{\rm rot}= {2 \pi R_{\rm sdB}}/{P}$ can be calculated. 
The radius of the primary is given by the relation 
$R = \sqrt{{M_{\rm sdB}G}/{g}}$. The measurement of the 
projected rotational velocity $v_{\rm rot}\,\sin\,i$ therefore allows us to 
constrain the inclination angle $i$. 
$M_{\rm sdB}$ remains the only free parameter. We varied $M_{\rm sdB}$, 
solved the mass function, and derived both the 
inclination angle and the companion mass. 
Because of $\sin{i} \leq 1$ a lower limit for the sdB mass is given by 
$M_{\rm sdB} \geq v_{\rm rotsini}^{2} P^{2}g/4 \pi^{2}G$ \citep{geier07,geier10b}.

We can constrain the mass of the sdB to be higher than $0.49\,M_{\rm \odot}$, and the mass of the WD companion to be higher than $0.74\,M_{\rm \odot}$ in this
way. 

A quantitative analysis of the light curve obtained with SOAR/Goodman fitting synthetic models is ongoing and necessary to constrain the parameters more accurately. 
However, the fact that CD$-$30\,11223 is eclipsing already tells us that the inclination must be high ($\simeq70-90^{\rm \circ}$). With all these parameters reasonably well  constrained we find that CD$-$30\,11223 has the smallest separation ($\simeq0.6\,R_{\rm \odot}$) of all known sdB binaries.

\section{Conclusions}

Due to the emission of gravitational waves, the binary is predicted to shrink until the sdB fills its Roche lobe and starts mass transfer in less then $50\,{\rm Myr}$. According to its atmospheric parameters, the sdB has just started the core helium-burning phase, which typically lasts for about $100\,{\rm Myr}$. Helium-rich material is therefore transferred to the WD companion, which according to its mass must consist of carbon and oxygen. 

This mass transfer is likely to be stable and will lead to the accumulation of a He-layer on top of the WD. After accreting $\simeq0.15\,M_{\rm \odot}$ He-burning is predicted to be ignited in this shell. This in turn can trigger the ignition of carbon in the core even if the WD-mass is significantly lower than the Chandrasekhar limit \citep{nomoto82,woosley86}. Hydrodynamic simulations predict the explosion of a CO-WD with a minimum mass of only $\simeq0.8\,M_{\rm \odot}$ triggered by the ignition of an He-shell of $\simeq0.1\,M_{\rm \odot}$ \citep{fink10}. These parameters closely match the properties of CD$-$30\,11223, which therefore becomes a viable candidate for SN\,Ia progenitor via this so-called double-detonation scenario.

%\bibliography{geier}

\begin{thebibliography}{}

\bibitem[Bloemen et al.(2011)]{bloemen11}
Bloemen, S., et al. 2011, MNRAS, 410, 1787
\bibitem[Fink et al.(2010)]{fink10} 
Fink, M., et al. 2010, A\&A, 514, 53
\bibitem[Geier et al.(2010a)]{geier10a}
Geier, S., Heber, U., Kupfer, T., \& Napiwotzki, R. 2010a, A\&A, 515, 37
\bibitem[Geier et al.(2010b)]{geier10b}
Geier, S., Heber, U., Podsiadlowski, Ph., et al. 2010b, A\&A, 519, 25
\bibitem[Geier et al.(2007)]{geier07}
Geier, S., Nesslinger, S., Heber, U., et al. 2007, A\&A, 464, 299
\bibitem[Geier et al.(2011a)]{geier11a}
Geier, S., Hirsch, H., Tillich, A., et al, 2011a, A\&A, 530, 28
\bibitem[Geier et al.(2011b)]{geier11b}
Geier, S., Maxted, P. F. L., Napiwotzki, R., et al. 2011b, A\&A, 526, 39
\bibitem[Geier et al.(2011c)]{geier11c}
Geier, S., Schaffenroth, V., Drechsel, H., et al. 2011c, ApJ, 731, L22
\bibitem[Han et al.(2002)]{han02}
Han Z., Podsiadlowski P., Maxted P. F. L., Marsh T. R., \& Ivanova N. 2002, MNRAS, 336, 449
\bibitem[Han et al.(2003)]{han03}
Han, Z., Podsiadlowski, P., Maxted, P. F. L., \& Marsh, T. R. 2003, MNRAS, 341, 669
\bibitem[Maxted et al.(2000)]{maxted00}
Maxted, P. F. L., Marsh, T. R., \& North, R. C. 2000, MNRAS, 317, L41
\bibitem[Maxted et al.(2001)]{maxted01}
Maxted, P. F. L., Heber, U., Marsh, T. R., \& North, R. C. 2001, MNRAS, 326, 139 
\bibitem[Napiwotzki et al.(2004)]{napiwotzki04}
Napiwotzki, R., Karl, C., Lisker, T., et al. 2004, Ap\&SS, 291, 321
\bibitem[Nomoto(1982)]{nomoto82}
Nomoto, K. 1982, ApJ 257, 780
\bibitem[Pollacco et al.(2006)]{pollacco06}
Pollacco, D. L., et al. 2006, PASP, 118, 1407
\bibitem[Vennes et al.(2011)]{vennes11}
Vennes, S., Kawka, A., \& N\'emeth, P. 2011, MNRAS, 410, 2095
\bibitem[Wang \& Han(2012)]{wang12}
Wang, B., \& Han, Z. 2012, NewAR, 56, 122
\bibitem[Woosley et al.(1986)]{woosley86} 
Woosley, S. E., Taam, R. E., \& Weaver, T. A. 1986, ApJ, 301, 601


\end{thebibliography}

\end{document}